**Drivers of academic engagement in public-private research collaboration: an empirical study**

Giovanni Abramo
*Laboratory for Studies in Research Evaluation*
*at the Institute for System Analysis and Computer Science (IASI-CNR)*
*National Research Council of Italy*
ADDRESS: Istituto di Analisi dei Sistemi e Informatica, Consiglio Nazionale delle Ricerche,
Via dei Taurini 19, 00185 Roma - ITALY
ORCID: 0000-0003-0731-3635 - giovanni.abramo@uniroma2.it

Ciriaco Andrea D'Angelo
*University of Rome "Tor Vergata" - Italy and*
*Laboratory for Studies in Research Evaluation (IASI-CNR)*
ADDRESS: Dipartimento di Ingegneria dell'Impresa, Università degli Studi di Roma "Tor Vergata", Via del Politecnico 1, 00133 Roma - ITALY
ORCID: 0000-0002-6977-6611 - dangelo@dii.uniroma2.it

**Abstract**
University-industry research collaboration is one of the major research policy priorities of advanced economies. In this study, we try to identify the main drivers that could influence the propensity of academics to engage in research collaborations with the private sector, in order to better inform policies and initiatives to foster such collaborations. At this purpose, we apply an inferential model to a dataset of 32,792 Italian professors in order to analyze the relative impact of individual and contextual factors affecting the propensity of academics to engage in collaboration with industry, at overall level and across disciplines. The outcomes reveal that the typical profile of the professor collaborating with industry is a male under age 40, full professor, very high performer, with highly diversified research, and who has a certain tradition in collaborating with industry. This professor is likely to be part of a staff used to collaborating with industry, in a small university, typically a polytechnic, located in the north of the country.

***<u>Declarations of interest: none</u>***

* *corresponding author*

## 1. Introduction

Increasingly, the capacity of a nation to produce wealth depends on investment in strengthening the so-called "triangle of knowledge": research, education and innovation. Many analyses show that competitive capacity is favored by policies that stimulate technological transfer and incentivize osmosis between the worlds of academic research and industry (Fan, Yang, & Chen, 2015; Shane, 2004; Thursby & Thursby, 2003; Mowery, Nelson, Sampat, & Ziedonis, 2001). Indeed, all industrially-developed nations have identified cross-sector co-operation as a major policy priority.

Scholars in the economics of innovation have extensively analyzed the development of public-private interaction. The relations between the two sectors take form through various modalities and can differ in their degree of formalization (Perkmann et al., 2013), with typical paths including joint research projects, the award of research contracts, licensing and award of know-how and patents, and through consulting, training services and professional mobility. The observation of such modalities, their empirical study and analysis of their underlying determinants, can provide useful cognitive bases for the policy maker and for research managers called to stimulate them.

One of the main modes of knowledge osmosis is through public-private research collaboration. The studies investigating this phenomenon have inquired into a range of subjects. One line of research concerns the identification of individual and institutional motivations, thereby permitting analysis of the drivers of collaboration. The determinants of variety and frequency of public-private interactions lie above all in the characteristics of the individual researcher, much more than those of their home organizations (D'Este & Patel, 2007), but these are also affected by the disincentive of transaction costs, increasing with greater cultural and motivational differences between partners (Belkhodja & Landry, 2005; Drejer & Jørgensen, 2005).

Within this research stream, we intend to add knowledge on the main drivers that could influence the propensity of academics to engage in research collaborations with the private sector. Previous studies investigating this phenomenon have generally resorted to surveys, which embed severe limits on the scale of observations. To overcome these, we adopt a bibliometric approach. Co-authorship of scientific publications can certify research collaborations, even though not all co-authored publications reveal a true research collaboration, and not all research collaborations bring to co-authored scientific publications (Katz & Martin, 1997). Further considerations are that industry relies heavily on collaboration with academia in publishing papers, but not in patenting, while academia rarely turns to industry in matters of research publication or patent applications (Huang, Yang, & Chen, 2015).

We will consider three sets of determinants of the intensity of university-industry collaboration at individual level.

Because social aspects are at play (Llopis, Sánchez-Barrioluengo, Olmos-Peñuela, & Castro-Martínez, 2018), we expect that individual traits (e.g. age, gender, academic rank, and research performance) will have an influence.

We also expect that contextual factors, particularly those of the nature, size and location of a university, could influence the frequency and scope of opportunities for academics to personally interact with colleagues in industry (Zhao, Broström, & Cai, 2020).

Finally, it is known that the research production function varies across scientific domains (Abramo, D'Angelo, & Murgia, 2013a), therefore we also expect significant discipline effects on the intensity of university-industry collaborations, certainly more noticeable in applied-science disciplines than theoretical ones. Our research questions can therefore be summarized as:

- What is the typical profile of academic professors who collaborate with the private sector?
- Which individual and contextual drivers have the greatest impact on the propensity for research collaboration with the private sector?
- How does the importance of such drivers vary across research fields?

To address these questions we adopt a statistical approach, applying an inferential model on a dataset of 32,792 professors working in Italian universities, in the hard sciences. This is the first study on individual-level collaboration based on bibliometric techniques, which allow large-scale investigations.

As noted, all industrially developed countries, in various ways, have enacted policy programs incentivizing university-industry collaboration. In recent years, Italy has followed the example of the UK research evaluation framework (REF), by enacting national research assessment exercises (VQRs) for evaluation of university "third mission" activities, which then support allocation of public funding.

By focusing on university researchers and the factors influencing their interactions with industry, we can improve our understanding about who in academia interacts with industry, and why (Bercovitz & Feldman, 2003). The empirical evidence from our study can serve decision makers engaged in formulating policies and incentive systems, intended to foster university-industry knowledge flows. Ex post, they can support assessment of the effectiveness of existing incentives, net of the determinants of individual propensity to engage in cross-sector collaborations.

The next section of this work provides a review of previous studies on the determinants of private-public research collaboration. Section three presents the dataset and describes the methodology adopted. The fourth section presents and comments the results of the analysis. The work concludes with a discussion of the findings and their implications for research policies.

## 2. Theoretical framework

Research collaboration between university and industry has positive effects in economic growth, social value, and competitiveness. Observing this, policy makers have increased their efforts to promote public-private interactions, acting on both sides of the partnerships through: i) research policies intended to promote the so-called third mission of universities (Iorio, Labory, & Rentocchini, 2016); ii) innovation policies that encourage companies to interact with public research organizations (Perkmann et al., 2013; Guimón & Paunov, 2019).

The underlying motivations for public-private co-operation are different for the two partners. Private enterprise is interested in collaborating with the public sector for access to skills, with which to create new knowledge for aims of commercial exploitation (Bekkers & Bodas Freitas, 2008; Perkmann et al., 2013). On the other side, academics benefit from accessing instrumental assets, but also gain direct economic and financial benefits of different kinds (Garcia, Araújo, Mascarini, Santos, & Costa, 2020). Cross-sector collaborations involve costs, however, so the propensity and intensity of public-private interactions depend on the balance between costs and benefits of such relationships.

Previous literature has focused almost entirely on observation of the phenomenon from the perspective of the academic researcher, generally through surveys on small national samples: 564 Chinese scientists in Zhao, Broström, and Cai (2020); 178 academics in Sri Lanka in Weerasinghe and Dedunu (2020); 1,295 researchers active at the Spanish National Research Council in Llopis, Sánchez-Barrioluengo, Olmos-Peñuela, and Castro-Martínez (2018); 4,400 Norwegian academics in Thune, Reymert, Gulbrandsen, and Aamodt (2016); 4337 UK engineering and physics scientists in D'Este and Patel (2007).

Such surveys indicate that propensity and extent of engagement by public researchers in joint research project with industry is influenced by several variables, some referring to the individual characteristics of the researcher, others to their context of operation.

### 2.1 Individual characteristics

Many studies reveal a correlation between the propensity to collaborate with companies and the career cycle of an academic, although the impact of age itself seems non-linear (Weerasinghe & Dedunu, 2020). One might think that because of their greater social capital, older researchers would have more intense and diversified collaborative activities (Bozeman & Corley, 2004). In fact, collaborations tend to grow most strongly early in the scientist's career, as they begin to establish a reputation, but seem to decline in the later stages, possibly as the initial incentives drop off (Bozeman & Gaughan, 2011; Ubfal & Maffioli, 2011). Younger academics are also pushed to collaborate to overcome disadvantages in availability of resources, and to demonstrate their capacity to activate and manage collaborations, considered essential to career progress (Bayer & Smart, 1991; Traoré & Landry, 1997).

Age and academic rank are strongly correlated, given the linkages between seniority and promotion mechanisms. However, in this type of study the effect of age should be separated from effect of researcher status. All others being equal, the academic rank of a scientist, attesting their authority in the community of reference, impacts on their ability to attract the interest of private companies (Abramo, D'Angelo, & Murgia, 2014; D'Este & Patel, 2007).

Gender, in general, also plays an important role in the propensity and intensity of collaborations, and in particular for cross-sector collaborations. The overall representation of women in research is increasing, but inequality remains. Mechanisms of gender homophily would clearly contribute to the greater difficulty of women in developing their social capital, and therefore accessing top positions, research funding and collaboration networks (Boschini & Siogren, 2007). Analyzing a large sample of UK physics and engineering scientists, Tartari and Salter (2015) argue that women academics engage less in collaboration activities with industry than male colleagues of similar status, as well as in different ways. Calvo, Fernández-López, and Rodeiro-Pazos (2019) trace the male/female gap to motivational reasons: surveying a sample of 420 research groups of eight regions of Spain, France and Portugal, they found that research groups led by women have lower probability of showing interest in R&D co-operation with firms.

The early behaviour of an individual academic in joint research with industry generates a strong imprint, leading to expectations of continuing knowledge transfer practices (Bercovitz & Feldman, 2003).

The research quality of the academic and their activity in technology transfer are found to be positively related (Mansfield, 1995; Mansfield & Lee, 1996). Generally, the link between research collaboration and performance is amply accepted in the literature (He, Geng, & Campbell-Hunt, 2009; Lee & Bozeman, 2005; Schartinger, Schibany, & Gassler, 2001). Although the causal nexus appears complex and variable (Abramo, D'Angelo, & Murgia, 2017), it is logical that there would be positive correlation between a professor's scientific performance and the intensity of their collaboration with private companies, who on the other side are searching top partners for their R&D projects (Balconi, & Laboranti, 2006). However, in a market strongly hampered by information asymmetry, companies could often base their decisions on geographical and social proximity more than on the standing of their potential partners (Abramo, D'Angelo, Di Costa, & Solazzi, 2011).

Regarding the empirical verification of a possible link between research collaboration and performance, many studies have relied simply on the aggregate profile of the department or institution as a proxy of the research quality of the individual scientists (Perkmann, King & Pavelin, 2011). The current study instead begins with the profile of the individual, including a variable that measures the level of specialization/diversification of the scientist's research activities, a characteristic not previously considered. Regarding the effect of this aspect on the propensity to collaborate with private companies, two opposing hypotheses seem possible: on the one hand, companies could be on the lookout for scientists highly specialized in their fields of interest; on the other hand, a varied and well-sorted scientific portfolio is more likely to intersect with some topic of interest to a private company, on which the academic can graft a joint research project. These are hypotheses that have never been proposed in the literature and therefore constitute one of the novelties of this study.

## 2.2 Contextual characteristics

Clearly, contextual characteristics will have an influence on the academic's propensity to engage in joint research with industry.

The university's statement of core mission has a strong effect on the research trajectories of its scientific laboratories. The influence of the "third mission" on the research agenda of an academic depends on the "commercial orientation" of their university (Di Gregorio & Shane, 2003). A university or department with a mission explicitly including collaboration with industry can certainly provide significant motivational stimuli for its individual researchers (Giuri, Munari, Scandura, & Toschi, 2019). The step of operating a technology transfer office is generally found to be a positive determinant of collaboration (Phan & Siegel, 2006), helping in particular to reduce "cognitive distance" between academics and industry (Muscio & Pozzali, 2013).

The intensity of overall activity in technology transfer, and of public-private research collaboration in particular, is certainly correlated with the demand for new knowledge expressed by companies in a given territory, meaning that the location of the researcher's home university also affects the possibilities of their collaboration with private colleagues (Berbegal-Mirabent, Sánchez García, & Ribeiro-Soriano, 2015).

It is plausible that the size of the university department would be related to the intensity of interactions with industry. The absence of a critical mass of staff or assets instrumental to the academic's specific research area could encourage them to seek collaborations with colleagues of other organizations (Schartinger, Schibany, & Gassler, 2001). Scarcity of institutional funding can push departments/groups to pursue an "income generation strategy" with respect to private sources (Giuri, Munari, Scandura, & Toschi, 2019). At the same time, an abundance of industry-sourced funding may signal an environment that favors institutional interaction with private companies (Schartinger, Schibany, & Gassler, 2001).

Finally, Tartari, Perkmann, & Salter (2014) argue that "academic scientists' industry engagement is influenced significantly by the behaviour of their peers, … through the mechanism of social comparison". The researcher's engagement will vary in the course of their career, but in any case will be conditioned by the choices of their reference group, both because they find inspiration there and because through conformity, they achieve social approval from their peers.

## 3. Data and methods

The field of observation consists of Italian university professors conducting research in the so-called hard sciences. We exclude the social sciences and arts & humanities because for these the coverage of bibliographic repertories is still insufficient for reliable representation of research output. In the Italian university system all professors are classified in one and only one field (named the scientific disciplinary sector, SDS, 370 in all). Fields are grouped into disciplines (named university disciplinary areas, UDAs, 14 in all).

The analysis dataset consists of all assistants, associate and full professors (34,410 in all), on staff for at least three years in the period 2013-2017, in 201 SDSs (of 10 UDAs), where publications in international journals serve as a reliable proxy for overall research output. The bibliometric dataset is extracted from the Italian Observatory of Public Research, a database developed and maintained by the present authors, and derived under license from the Clarivate Analytics' Web of science (WoS). Beginning from the raw data of the WoS, and applying a complex algorithm to reconcile the authors' affiliations and disambiguate their true identity, each publication (article, article review and conference proceeding) is attributed to the university scientist or scientists that produced it (D'Angelo, Giuffrida, & Abramo, 2011).[1]

The result of the algorithm application is as follows. 1618 (4.7%) professors are unproductive in the five years under observation. Of the 32,792 productive professors, the analysis of co-authorship of publications shows that 9005 (27.5%) professors co-authored at least one publication with industry (Table 1). At the UDA level, Industrial and information engineering has the highest share of productive professors collaborating with industry (47.5%), followed by Chemistry (37.4%). Psychology and Mathematics and computer science are the UDAs with lowest percentages, respectively 5.5% and 11.0%. It should be noted that D'Este and Patel (2007) found that over 40% of UK physics and engineering researchers were involved in at least one industry interaction over 2002-2003: the comparable statistic is obtained from Table 1 by collapsing UDAs 2 and 9.

[1] The harmonic average of precision and recall (F-measure) of authorships, as disambiguated by the algorithm, is around 97% (2% margin of error, 98% confidence interval).

*Table 1: Dataset of the analysis*

| UDA | SDS | No. of Professors | With at least one publication | With at least one publication with industry |
|---|---|---|---|---|
| 1 - Mathematics and computer science | 10 | 3083 | 2797 (90.7%) | 308 (11.0%) |
| 2 - Physics | 8 | 2193 | 2120 (96.7%) | 491 (23.2%) |
| 3 - Chemistry | 11 | 2844 | 2803 (98.6%) | 1047 (37.4%) |
| 4 - Earth sciences | 12 | 1031 | 996 (96.6%) | 317 (31.8%) |
| 5 - Biology | 19 | 4729 | 4599 (97.3%) | 1167 (25.4%) |
| 6 - Medicine | 50 | 9411 | 8876 (94.3%) | 2029 (22.9%) |
| 7 - Agricultural and veterinary sciences | 30 | 2978 | 2860 (96.0%) | 791 (27.7%) |
| 8 - Civil engineering | 9 | 1505 | 1402 (93.2%) | 374 (26.7%) |
| 9 - Industrial and information engineering | 42 | 5248 | 5081 (96.8%) | 2412 (47.5%) |
| 10 - Psychology | 10 | 1388 | 1258 (90.6%) | 69 (5.5%) |
| Total | 201 | 34410 | 32792 (95.3%) | 9005 (27.5%) |

To answer the research questions, we will use a logit regression with individual professors as unit of analysis. The dependent variable (y) is a dummy assuming: 1, if professor *i* co-authored at least one publication with industry; 0, otherwise. As discussed in Section 2, the literature suggests a number of drivers (covariates) that are likely to affect the propensity of professors to engage in research collaboration with industry. We consider the following covariates, grouped in two clusters.

Individual covariates

- Gender ($X_1$), specified by a dummy variable (1 for female; 0 for male);
- Age ($X_{2-5}$), specified with 5 classes, through 4 dummies (baseline "Under 40");
- Academic rank ($X_{6-7}$), specified by 2 dummies (baseline "Assistant professors");
- Level of specialization of the professor's scientific activity, specified by 2 dummies:
  - ✓ "Highly diversified" ($X_8$), 1 if the papers falling in the prevalent subject category of the professor are less than 40% of total publications; 0, otherwise;[2]
  - ✓ "Highly specialized" ($X_9$), 1 if the papers falling in the prevalent subject category of the professor are more than 75% of total publications; 0, otherwise;[2]
- Past behaviour - previous collaborations ($X_{10}$), specified by a dummy variable (1 if the professor co-authored at least one publication with industry in 2010-2012; 0, otherwise);
- Total publications authored by the professor in the period under observation ($X_{11}$);
- Research performance as measured by the FSS (fractional scientific strength) indicator:[3]
  - ✓ $X_{12}$, FSS of the professor, rescaled to the field average;
  - ✓ $X_{13}$, 1 if the professor belongs to the top 20% in his/her field by FSS; 0, otherwise;

[2] The chosen thresholds (40% for $X_8$, and 75% for $X_9$) allow for equal partitions of the dataset (one third of of highly diversified professors, and one third of highly specialized ones).

[3] For details on the conceptualization of this indicator as a measure of scientific productivity, also for operational definition, see Abramo and D'Angelo (2014).

Contextual covariates

- Environment - Peers behaviour ($X_{14}$), specified by a dummy variable (1 in case of a colleague in the same university and SDS of the professor, co-authoring publications with industry; 0, otherwise);
- Institutional control ($X_{15}$), specified by a dummy variable (0, for public universities; 1, for private ones);
- University scope ($X_{16}$), specified by a dummy variable (1, for "Polytechnics" and "Schools for Advanced Studies, SS"; 0, otherwise);[4]
- University size in the UDA of the professor ($X_{17-18}$), specified with 3 classes through 2 dummies ("Large", for universities with a research staff in the UDA of the professor, above 80 percentile in the national ranking; "Medium", with a research staff between 50 and 80 percentile; baseline "Small");
- University location ($X_{19-22}$), specified with 5 macro-areas, by 4 dummies, baseline "Islands").

To control for the research discipline effects, we also consider other 9 dummies related to the ten UDAs under observation.

All variables are measured at 31/12/2012, i.e. the beginning of the period under observation.

Through the variables $X_{15}$ (Institutional control) and $X_{16}$ (University scope), the authors hope to intercept, although indirectly, the effect known in the literature as the "commercial orientation" of a university. Differently from the literature, our model does not embed the effect of the presence of a technology transfer office, as all Italian universities are required to have one by law. It is possible that the attention and resources devoted to them vary across universities, but we are unable to measure them.

## 4. Results

Table 2 reports the typical profile of the university professor collaborating with private companies, identified by the concentration index on each individual and contextual trait.[5] The academic is male, under age 40, a full professor, conducting research in Industrial and information engineering, with high scientific productivity (top 20%) and highly diversified research activity, as well as previous experience in cross-sector collaborations. This professor operates within a group of peers who likewise collaborate with industry, and belongs to a large public university, typically a polytechnic or SSs located in northwestern Italy.

[4] The Italian Minister of University and Research (MUR) recognizes a total of 96 universities as having the legal authority to issue degrees. Of these, 29 are small, private, special-focus universities, of which 13 offer only e-learning; 67 are public and generally multi-disciplinary universities; three are Polytechnics and six are *Scuole Superiori* (Schools for Advanced Studies), devoted to highly accomplished students, with very small faculties and tightly limited enrolment. In the overall system, 94.9% of faculty are employed in public universities (0.5% in *Scuole Superiori*).

[5] The concentration index is the ratio of two ratios. Example: for the group variable "gender", the prevailing trait "male" shows a concentration index of 1.069, since males compose 70.81% of total researchers co-authoring publications with industry, and 66.25% of the total population, therefore 70.81/66.25=1.069.

*Table 2: Profiling of the Italian academic professor co-authoring publications with industry*

| Group variable | Prevailing trait | Concentration index | Pearson chi-squared | p-value |
|---|---|---|---|---|
| Gender | Male | 1.069 | 115.0055 | 0.000 |
| Age | Under 40 | 1.123 | 159.2152 | 0.000 |
| Academic rank | Full professor | 1.134 | 79.4962 | 0.000 |
| UDA | 9 - Industrial and inform. engineering | 1.725 | 2.0e+03 | 0.000 |
| Scientific activity | Highly diversified | 1.128 | 146.4991 | 0.000 |
| Past behaviour | Previous collaboration | 2.275 | 2.6e+03 | 0.000 |
| Research performance (FSS quintile) | Top 20% | 1.434 | 1.2e+03 | 0.000 |
| Environment | Peers collaborating with industry | 1.233 | 1.5e+03 | 0.000 |
| University type | Public | 1.010 | 33.9566 | 0.000 |
| | Polytechnic or SS | 1.667 | 341.9875 | 0.000 |
| University size | Large | 1.021 | 10.0636 | 0.007 |
| University location | Northwest | 1.163 | 188.1679 | 0.000 |

However, the previous descriptive analysis does not take into account the simultaneous effect of all covariates on the independent variable. Therefore, we conduct an inferential analysis using a logit regression model, as illustrated in the previous section. Table 3 shows the average values of the model variables, at an overall level and by UDA. Table 4 shows the correlation matrix of the variables, at an overall level.

Some significant correlations between pairs of variables (shaded values in Table 4) deserve comment:

Y vs $X_{10}$ (Previous collaborations) – As could be expected, the probability of a professor establishing scientific collaborations with the private sectors is strongly influenced by the fact that they have done so before.

$X_5$ (Age-over_60) vs $X_7$ (Full prof.) - It is highly reasonable that age and academic rank would be strongly correlated, and that the attainment of the top level of the academic career takes place on average in old age since, as already mentioned, promotion mechanisms are often linked to seniority (very true in Italy).

$X_{11}$ (Tot. publications) vs $X_{12}$ (rescaled FSS) vs $X_{13}$ (Top 20%) - Output, a fundamental dimension of productivity, is incorporated in FSS together with the impact of each publication and the fractional contribution attributable to each co-author. This trio of variables has the highest correlation index, which calls for checking possible multicollinearity. In reality, as we will see, the co-presence of these covariates does not disturb the model. Rather, $X_{11}$ acts as an exposure variable, since the publications in collaboration with private companies (Y) are a subset of the total ones ($X_{11}$): the possible exclusion of this last variable would cause problems of convergence of logit regression.

$X_{14}$ (Peers behaviour) vs $X_{18}$ (University size - Large) - Expected correlation, since the probability of a professor having colleagues in the same field, with at least one publication in collaboration with companies, depends on the size of the overall faculty research staff.

$X_{16}$ (University type: Polytechnic or SS) vs $X_{22}$ (University location: Northwest) – Reflects a feature typical of the Italian academic system, descending from the facts that the Northwest is the most industrialized area and hosts two of the country's three polytechnics.

***Table 3: Average values of the regression model variables by UDA***

| Variable group | | UDA§ | 1 | 2 | 3 | 4 | 5 | 6 | 7 | 8 | 9 | 10 | Total |
|---|---|---|---|---|---|---|---|---|---|---|---|---|---|
| Response | Y | Co-authorships with industry | 0.112 | 0.231 | 0.374 | 0.316 | 0.253 | 0.228 | 0.278 | 0.268 | 0.474 | 0.058 | 0.275 |
| Gender | $X_1$ | Female | 0.329 | 0.202 | 0.458 | 0.295 | 0.528 | 0.317 | 0.378 | 0.213 | 0.162 | 0.554 | 0.337 |
| Age | $X_2$ | 40-45 | 0.223 | 0.182 | 0.211 | 0.169 | 0.182 | 0.128 | 0.180 | 0.230 | 0.229 | 0.254 | 0.185 |
| | $X_3$ | 46-52 | 0.252 | 0.278 | 0.252 | 0.333 | 0.263 | 0.213 | 0.297 | 0.270 | 0.243 | 0.191 | 0.248 |
| | $X_4$ | 53-60 | 0.184 | 0.199 | 0.202 | 0.243 | 0.261 | 0.359 | 0.268 | 0.158 | 0.153 | 0.166 | 0.247 |
| | $X_5$ | Over 60 | 0.107 | 0.160 | 0.130 | 0.119 | 0.141 | 0.188 | 0.089 | 0.121 | 0.094 | 0.115 | 0.136 |
| Academic rank | $X_6$ | Associate prof. | 0.284 | 0.318 | 0.293 | 0.309 | 0.254 | 0.278 | 0.280 | 0.276 | 0.281 | 0.243 | 0.279 |
| | $X_7$ | Full prof. | 0.270 | 0.211 | 0.194 | 0.198 | 0.204 | 0.197 | 0.226 | 0.263 | 0.261 | 0.212 | 0.221 |
| Scientific activity | $X_8$ | Highly diversified | 0.102 | 0.193 | 0.297 | 0.110 | 0.405 | 0.265 | 0.202 | 0.108 | 0.151 | 0.342 | 0.237 |
| | $X_9$ | Highly specialized | 0.400 | 0.271 | 0.098 | 0.121 | 0.073 | 0.173 | 0.188 | 0.248 | 0.253 | 0.138 | 0.192 |
| Past behaviour | $X_{10}$ | Previous collaborations | 0.035 | 0.099 | 0.185 | 0.137 | 0.113 | 0.103 | 0.103 | 0.083 | 0.184 | 0.016 | 0.115 |
| Research performance | $X_{11}$ | Tot. publications | 11.1 | 46.9 | 21.8 | 14.2 | 15.7 | 22.9 | 14.3 | 14.6 | 23.4 | 12.6 | 20.6 |
| | $X_{12}$ | FSS | 0.946 | 0.991 | 0.996 | 0.993 | 0.993 | 0.984 | 0.984 | 0.968 | 0.981 | 0.960 | 0.982 |
| | $X_{13}$ | Top 20% | 0.222 | 0.208 | 0.205 | 0.213 | 0.208 | 0.215 | 0.212 | 0.217 | 0.209 | 0.224 | 0.212 |
| Environment | $X_{14}$ | Peers behaviour | 0.507 | 0.766 | 0.860 | 0.533 | 0.737 | 0.678 | 0.660 | 0.603 | 0.816 | 0.273 | 0.690 |
| University type | $X_{15}$ | Private | 0.021 | 0.016 | 0.004 | 0.000 | 0.025 | 0.073 | 0.027 | 0.023 | 0.022 | 0.128 | 0.038 |
| | $X_{16}$ | Polytechnic or SS | 0.067 | 0.076 | 0.020 | 0.021 | 0.007 | 0.000 | 0.004 | 0.183 | 0.233 | 0.002 | 0.058 |
| University size | $X_{17}$ | Medium | 0.335 | 0.368 | 0.360 | 0.398 | 0.335 | 0.344 | 0.296 | 0.337 | 0.298 | 0.247 | 0.331 |
| | $X_{18}$ | Large | 0.570 | 0.523 | 0.546 | 0.451 | 0.555 | 0.593 | 0.677 | 0.534 | 0.641 | 0.668 | 0.588 |
| University location | $X_{19}$ | South | 0.207 | 0.199 | 0.198 | 0.226 | 0.197 | 0.177 | 0.246 | 0.277 | 0.199 | 0.131 | 0.199 |
| | $X_{20}$ | Center | 0.238 | 0.254 | 0.225 | 0.249 | 0.266 | 0.285 | 0.209 | 0.213 | 0.241 | 0.232 | 0.251 |
| | $X_{21}$ | Northeast | 0.208 | 0.219 | 0.230 | 0.212 | 0.187 | 0.164 | 0.223 | 0.166 | 0.188 | 0.281 | 0.195 |
| | $X_{22}$ | Northwest | 0.263 | 0.233 | 0.212 | 0.204 | 0.229 | 0.252 | 0.182 | 0.239 | 0.293 | 0.256 | 0.243 |
| | | Obs. | 2793 | 2118 | 2808 | 996 | 4605 | 8871 | 2857 | 1402 | 5081 | 1261 | 32792 |

§ *1, Mathematics and computer science; 2, Physics; 3, Chemistry; 4, Earth sciences; 5, Biology; 6, Medicine; 7, Agriculture and veterinary sciences; 8, Civil engineering; 9, Industrial and information engineering; 10, Psychology*

***Table 4: Correlation matrix of the regression model variables***

| | Y | $X_1$ | $X_2$ | $X_3$ | $X_4$ | $X_5$ | $X_6$ | $X_7$ | $X_8$ | $X_9$ | $X_{10}$ | $X_{11}$ | $X_{12}$ | $X_{13}$ | $X_{14}$ | $X_{15}$ | $X_{16}$ | $X_{17}$ | $X_{18}$ | $X_{19}$ | $X_{20}$ | $X_{21}$ |
|---|---|---|---|---|---|---|---|---|---|---|---|---|---|---|---|---|---|---|---|---|---|---|
| Y | 1 | | | | | | | | | | | | | | | | | | | | | |
| $X_1$ | -0.059 | 1 | | | | | | | | | | | | | | | | | | | | |
| $X_2$ | 0.020 | 0.069 | 1 | | | | | | | | | | | | | | | | | | | |
| $X_3$ | 0.023 | 0.008 | -0.274 | 1 | | | | | | | | | | | | | | | | | | |
| $X_4$ | -0.031 | -0.014 | -0.273 | -0.329 | 1 | | | | | | | | | | | | | | | | | |
| $X_5$ | -0.053 | -0.123 | -0.190 | -0.228 | -0.228 | 1 | | | | | | | | | | | | | | | | |
| $X_6$ | 0.006 | -0.033 | -0.049 | 0.177 | 0.097 | -0.001 | 1 | | | | | | | | | | | | | | | |
| $X_7$ | 0.043 | -0.191 | -0.219 | -0.075 | 0.164 | 0.419 | -0.335 | 1 | | | | | | | | | | | | | | |
| $X_8$ | 0.044 | 0.083 | -0.005 | 0.024 | 0.012 | -0.027 | -0.002 | -0.027 | 1 | | | | | | | | | | | | | |
| $X_9$ | -0.060 | -0.052 | -0.015 | -0.011 | -0.005 | 0.037 | 0.001 | 0.019 | -0.272 | 1 | | | | | | | | | | | | |
| $X_{10}$ | 0.284 | -0.040 | 0.008 | 0.028 | 0.002 | 0.005 | 0.021 | 0.080 | 0.031 | -0.035 | 1 | | | | | | | | | | | |
| $X_{11}$ | 0.211 | -0.115 | 0.017 | 0.015 | -0.025 | -0.041 | 0.002 | 0.106 | -0.025 | 0.032 | 0.158 | 1 | | | | | | | | | | |
| $X_{12}$ | 0.151 | -0.091 | 0.033 | 0.002 | -0.064 | -0.073 | -0.012 | 0.060 | -0.004 | -0.021 | 0.103 | 0.521 | 1 | | | | | | | | | |
| $X_{13}$ | 0.139 | -0.085 | 0.028 | -0.007 | -0.058 | -0.072 | -0.004 | 0.047 | -0.007 | -0.007 | 0.081 | 0.425 | 0.658 | 1 | | | | | | | | |
| $X_{14}$ | 0.214 | 0.003 | -0.015 | 0.000 | -0.002 | 0.014 | -0.003 | -0.005 | 0.020 | -0.050 | 0.113 | 0.110 | 0.031 | 0.020 | 1 | | | | | | | |
| $X_{15}$ | -0.032 | -0.003 | -0.012 | -0.012 | 0.011 | -0.019 | -0.030 | -0.034 | 0.002 | -0.013 | -0.018 | 0.000 | 0.016 | 0.013 | -0.060 | 1 | | | | | | |
| $X_{16}$ | 0.102 | -0.067 | 0.034 | -0.002 | -0.058 | -0.030 | 0.002 | 0.022 | -0.044 | 0.034 | 0.056 | 0.031 | 0.012 | 0.018 | 0.106 | -0.049 | 1 | | | | | |
| $X_{17}$ | -0.008 | -0.013 | 0.002 | 0.008 | 0.003 | 0.002 | 0.005 | 0.010 | 0.000 | 0.002 | -0.014 | -0.015 | -0.009 | -0.006 | -0.144 | -0.058 | -0.128 | 1 | | | | |
| $X_{18}$ | 0.015 | 0.013 | -0.002 | -0.014 | 0.009 | 0.022 | 0.001 | -0.012 | -0.007 | 0.007 | 0.021 | 0.012 | 0.007 | 0.007 | 0.247 | -0.052 | 0.107 | -0.841 | 1 | | | |
| $X_{19}$ | -0.020 | -0.005 | 0.015 | 0.007 | -0.021 | -0.015 | -0.004 | 0.007 | 0.011 | -0.009 | -0.025 | -0.014 | -0.008 | -0.009 | -0.083 | -0.094 | -0.056 | 0.015 | -0.083 | 1 | | |
| $X_{20}$ | -0.007 | 0.003 | -0.026 | -0.001 | 0.055 | 0.016 | 0.005 | -0.007 | -0.003 | 0.012 | -0.010 | -0.004 | -0.019 | -0.020 | 0.024 | -0.035 | -0.109 | -0.048 | 0.047 | -0.289 | 1 | |
| $X_{21}$ | 0.011 | -0.017 | -0.006 | 0.006 | -0.008 | -0.007 | 0.016 | 0.005 | -0.020 | 0.013 | -0.001 | 0.019 | 0.033 | 0.032 | -0.004 | -0.074 | -0.103 | 0.080 | -0.058 | -0.247 | -0.286 | 1 |
| $X_{22}$ | 0.057 | 0.008 | 0.016 | 0.002 | -0.038 | -0.008 | -0.007 | -0.002 | -0.012 | 0.002 | 0.052 | 0.024 | 0.018 | 0.010 | 0.102 | 0.224 | 0.322 | -0.087 | 0.102 | -0.283 | -0.328 | -0.280 |

*Y, Co-authorships with industry; $X_1$, Female; $X_2$, 40-45; $X_3$, 46-52; $X_4$, 53-60; $X_5$, Over 60; $X_6$, Associate prof.; $X_7$, Full prof.; $X_8$, Highly diversified; $X_9$, Highly specialized; $X_{10}$, Previous collaborations; $X_{11}$, Tot. publications; $X_{12}$, rescaled FSS; $X_{13}$, top20% by FSS; $X_{14}$, Peers behaviour; $X_{15}$, Private; $X_{16}$, Politechnic or SS; $X_{17}$, Medium; $X_{18}$, Large; $X_{19}$, South; $X_{20}$, Center; $X_{21}$, Northeast; $X_{22}$, Northwest*

Table 5 presents the results of the logit regression. For simpler representation, the coefficients for UDA dummies are omitted. The in-depth analysis by UDA is reported at Table 6.

The model estimation appears very satisfactory. The mean VIF is 2.45, with maximum (6.99) for the covariate University size - Large, which excludes the presence of multicollinearity that could disturb the estimation of the coefficients. The value of under ROC area (AUC) is 0.77, which indicates good ability of the model to correctly classify professors, discriminating the propensity to collaborate with companies[6] in function of individual traits and context.

The estimated coefficients of the regression model are expressed in terms of odds ratios: the reference value is equal to one and indicates that the variable considered has no effect on the Y, i.e. on the probability that a professor has or has not collaborated with private companies. For values above one, the variable instead has a positive marginal effect, and vice versa. The data of column 7, Table 5 indicate that all the covariates have a statistically significant effect.

Among the individual characteristics, gender has a non-marginal effect: women show a propensity to collaborate with private companies 8.7% less than men. This data confirms the indications of previous studies on the subject, in particular Weerasinghe and Dedunu (2020), Tartari and Salter (2015), and Calvo, Fernández-López, and Rodeiro-Pazos (2019). The last authors found that "research groups led by men have around 10% higher probability of showing interest in R&D co-operation with firms", a datum absolutely similar to that obtained here despite obvious methodological differences.

Age seems to have a systematically negative impact on the response variable of the proposed model. The odds ratios detected for the four age classes considered are all lower than 1 and decreasing. In particular, compared to an under-40 professor, the probability of an over-60 collaborating in joint publication with companies is practically halved. This further confirms D'Este and Patel (2007): "the sign on the age variable is negative and the impact is significant […] suggesting that the younger the researcher the higher the probability of engaging in a greater variety of interactions, and also of engaging more frequently across a wider range of interactions". In contrast, Weerasinghe and Dedunu (2020) report that impact of age is non-linear, but this result may have been influenced by the particularly small sample size (178 academics of Sri Lankan state universities).

Academic rank shows a positive effect on the propensity for research collaboration with industry: full and associate professors show a higher probability compared to assistant professors, respectively by 39% and 18%, confirming the summary conclusion of Perkmann et al. (2013), who refer to a large number of previous works reporting a positive effect of professorial tenure on collaboration activities with industry.

Regarding the level of specialization/diversification, the estimation of the model returns a very interesting result: an academic professor with a highly diversified scientific profile has a 31.2% higher probability of collaborating with companies than an "intermediate" profile. Conversely, a highly specialized profile has a 27.2 percent lower probability of collaborating with companies. In this case, there is no benchmark in the literature to report this result, therefore constituting an effective advancement of knowledge on the subject.

[6] The AUC analysis evaluates a classifier's ability to discern between true positives and false positives. In our case, the AUC value, between 0 and 1, is equivalent to the probability that the result of the logit classifier applied to a researcher randomly extracted from the group of those who collaborated with industry is higher than that obtained by applying it to a researcher randomly extracted from the group of those who did not collaborate (Bowyer, Kranenburg, & Dougherty, 2001).

The data in column 4 Table 5 confirm the indication of the correlation analysis, on the fact that the main driver of an academic's public-private research collaboration is their track record ($X_{10}$). Compared to professors without, those with a previous collaboration have 300% greater probability of continuing. The data confirms the conclusions of D'Este and Patel (2007) that "those researchers with a record of past interaction are more likely to be involved [...] with industry."

A particularly interesting result concerns the effect of research performance: net of what can be considered an exposure variable (the number of total publications), the OR of the Top 20% variable (1.228) indicates that a standing in the top performance quintile increases the probability of engaging with industry by 23%. Contrarily, the coefficient relative to FSS is quite low (1.055), although positive and statistically significant. The latter result is in line with Abramo, D'Angelo, Di Costa, and Solazzi (2011), who interpret it in the light of inefficiency in selection of academic collaborators by industry, due to the evident information asymmetry on the demand/supply sides of new knowledge. Blumenthal, Campbell, Causino, and Louis (1996) likewise failed to find any clear relationship between academic productivity and cross-sector collaboration activities.

The contextual characteristics all show significant effects under equal conditions. Apart from past experience, the presence of colleagues from the same field, actively collaborating with companies, is the most important driver for an academic (odds ratio 2.714), confirming the indication of Tartari, Perkmann, and Salter (2014).

In Italy, the public character of the university also matters. In private universities, all other things being equal, the propensity to collaborate is lower (odds ratio 0.785). We have found no studies on the subject in the literature to compare our findings with. Professors of polytechnics and SSs show a higher propensity (+14.6%) compared to those of generalist universities. Again, comparison with prior knowledge is unavailable, except for a similar result achieved by D'Este and Patel (2007) in the UK, but referring only to institutions formerly tagged as polytechnics (i.e. higher education institutions, upgraded to university status in 1992). On the other hand, such results are no surprise, considering that research groups concentrated on the technical sciences are more likely to attract industry attention.

Regarding the size of the research staff, the coefficient of both variables considered (0.870 and 0.725) is significant and well below one, indicating that as the size increases, the propensity to co-publication with industry decreases. The results seem new with respect to knowledge on the subject. Schartinger, Schibany, and Gassler (2001) had expected a U-shaped curve of size effects on interactions, while for D'Este and Patel (2007) departmental variables (including the size of the relative research staff) "lose their significance once individual characteristics are introduced."

Finally, confirming the state of the art on the subject, location also matters: the coefficients of the four dummies considered indicate an increase in the propensity to collaborate with latitude (i.e. towards industrial territories of northern Italy). Compared to academic professors working in the two island regions (Sicily and Sardinia), all other things being equal, those of the south show a propensity greater by 24.2%; those of the center, 35%; those of the northeast, 39.3%; and finally those of the northwest, 48.2%. This pattern confirms that for university professors, interaction with companies depends on the territorial level of concentration of industrial activities in general and, in particular, of knowledge-intensive industry.

*Table 5: The main drivers of the propensity to collaborate with industry by Italian professors. Logit regression, dependent variable: 1 in case of publications in co-authorship with industry; 0, otherwise*

| | | | Odd ratio | Std Err. | z | P>z | [95% Conf. Interval] | |
|---|---|---|---|---|---|---|---|---|
| Variable group | | Const. | 0.053 | 0.005 | -29.6 | 0.000 | 0.044 | 0.064 |
| Gender | $X_1$ | Female | 0.913 | 0.029 | -2.88 | 0.004 | 0.858 | 0.971 |
| Age | $X_2$ | 40-45 | 0.899 | 0.041 | -2.35 | 0.019 | 0.822 | 0.983 |
| | $X_3$ | 46-52 | 0.824 | 0.039 | -4.06 | 0.000 | 0.751 | 0.905 |
| | $X_4$ | 53-60 | 0.698 | 0.037 | -6.78 | 0.000 | 0.629 | 0.775 |
| | $X_5$ | Over 60 | 0.526 | 0.035 | -9.62 | 0.000 | 0.461 | 0.599 |
| Academic rank | $X_6$ | Associate prof. | 1.175 | 0.045 | 4.2 | 0.000 | 1.090 | 1.268 |
| | $X_7$ | Full prof. | 1.387 | 0.067 | 6.76 | 0.000 | 1.262 | 1.526 |
| Scientific activity | $X_8$ | Highly diversified | 1.312 | 0.044 | 8.15 | 0.000 | 1.229 | 1.401 |
| | $X_9$ | Highly specialized | 0.728 | 0.029 | -7.98 | 0.000 | 0.673 | 0.787 |
| Past behaviour | $X_{10}$ | Previous collab. | 3.965 | 0.156 | 34.94 | 0.000 | 3.670 | 4.284 |
| Research performance | $X_{11}$ | Tot. publications | 1.012 | 0.001 | 17.05 | 0.000 | 1.010 | 1.013 |
| | $X_{12}$ | FSS | 1.055 | 0.016 | 3.59 | 0.000 | 1.025 | 1.087 |
| | $X_{13}$ | Top 20% | 1.228 | 0.056 | 4.52 | 0.000 | 1.123 | 1.342 |
| Environment | $X_{14}$ | Peers collaborating | 2.714 | 0.101 | 26.94 | 0.000 | 2.524 | 2.918 |
| University type | $X_{15}$ | Private | 0.785 | 0.066 | -2.88 | 0.004 | 0.666 | 0.925 |
| | $X_{16}$ | Polytechnic or SS | 1.146 | 0.072 | 2.17 | 0.030 | 1.013 | 1.296 |
| University size | $X_{17}$ | Medium | 0.870 | 0.050 | -2.4 | 0.016 | 0.776 | 0.975 |
| | $X_{18}$ | Large | 0.725 | 0.041 | -5.65 | 0.000 | 0.649 | 0.811 |
| University location | $X_{19}$ | South | 1.242 | 0.069 | 3.91 | 0.000 | 1.114 | 1.384 |
| | $X_{20}$ | Center | 1.350 | 0.072 | 5.65 | 0.000 | 1.216 | 1.498 |
| | $X_{21}$ | Northeast | 1.393 | 0.076 | 6.05 | 0.000 | 1.251 | 1.551 |
| | $X_{22}$ | Northwest | 1.482 | 0.082 | 7.15 | 0.000 | 1.331 | 1.652 |

Number of obs. = 32792
LR chi2(28) = 6158.1; Prob > chi2 = 0.0000
Log likelihood = -16010.5; Pseudo $R^2$ = 0.1613

The analysis was repeated for each UDA for purposes of detecting the importance of individual and contextual drivers across fields. Table 6 reports the results of the logit regression on the professors of each discipline. Although reducing the number of observations brings loss of significance for many coefficients, it remains evident that there is strong heterogeneity among the disciplines. Medicine (UDA 6), the area with the most observations (over 8,800), then also has the highest number of significant coefficients, whose values confirm the effects of covariates detected at overall level. Gender and age seem to lose consequence, except in Mathematics and computer science (UDA 1), where the effect of these personal traits seems even more pronounced than at overall level. In this same UDA, the odds ratio for “highly diversified” scientific activity (2.212) is very high, vs very low for the opposite character (highly specialized, 0.351). Although less pronounced than in UDA 1, this pair of variables shows the same type of effect in all other UDAs except 7, 8, 9.

Among individual traits, “Previous collaborations” remains the driver with the greatest weight, always significant, with minimum odds ratio in UDA 4 (Earth sciences, 2.320). Productivity (FSS) shows significant impact in all areas but Mathematics, Earth sciences, and Psychology (respectively UDAs 1,4, 10), but of opposite sign to that seen at the overall level (i.e. odds ratios always lower than one), with the sole exception of Physics (1.144).

Concerning the contextual effect of university size, the coefficients confirm, at least at sign level, what is observed at overall level. An exception is UDA 9 (Industrial and information engineering), where the marginal effect of size is positive, especially comparing between “small” and “medium” research faculties. The effect of the geographic localization variable seems heterogeneous and does not always confirming the primacy of the northwest among the different areas. In Mathematics and computer science no effect is detectable. Two other UDAs (Physics, Psychology) seem to present the rare case of the Italian islands offering greater opportunities for academic-industrial interaction than the rest of the country.

*Table 6: The main drivers of the propensity to collaborate with industry by Italian academic professors. Logit regression by UDA*

| | | UDA§ | 1 | 2 | 3 | 4 | 5 | 6 | 7 | 8 | 9 | 10 |
|---|---|---|---|---|---|---|---|---|---|---|---|---|
| Variable group | | Const | 0.087*** | 0.162*** | 0.103*** | 0.075*** | 0.076*** | 0.074*** | 0.057*** | 0.06*** | 0.083*** | 0.071*** |
| Gender | $X_1$ | Female | 0.584*** | 0.897 | 1.003 | 0.976 | 0.948 | 1.018 | 0.961 | 0.913 | 0.986 | 0.853 |
| Age | $X_2$ | 40-45 | 0.606** | 1.09 | 0.803 | 1.212 | 1.078 | 0.921 | 0.746* | 0.972 | 0.985 | 1.371 |
| | $X_3$ | 46-52 | 0.637** | 0.988 | 0.823 | 1.695** | 1.317** | 0.795** | 0.776 | 0.636* | 0.921 | 0.43* |
| | $X_4$ | 53-60 | 0.417*** | 0.743 | 0.783 | 1.103 | 1.065 | 0.761*** | 0.786 | 0.643 | 0.973 | 0.52 |
| | $X_5$ | Over 60 | 0.355*** | 0.562** | 0.658* | 1.25 | 0.888 | 0.602*** | 0.737 | 0.483** | 0.735* | 0.401 |
| Academic rank | $X_6$ | Associate prof. | 1.111 | 1.188 | 0.918 | 0.983 | 0.994 | 1.274*** | 0.949 | 1.42* | 0.912 | 2.754*** |
| | $X_7$ | Full prof. | 1.316 | 1.888*** | 0.856 | 0.957 | 0.872 | 1.365*** | 0.985 | 1.683** | 0.873 | 1.456 |
| Scientific activity | $X_8$ | Highly diversified | 2.212*** | 1.81*** | 1.271** | 1.709** | 1.133 | 1.307*** | 1.082 | 0.703 | 1.099 | 1.644* |
| | $X_9$ | Highly specialized | 0.351*** | 0.433*** | 0.879 | 0.499** | 0.55*** | 0.692*** | 1.193 | 1.07 | 1 | 0.507 |
| Past behaviour | $X_{10}$ | Previous collaborations | 5.295*** | 3.097*** | 4.079*** | 2.320*** | 3.818*** | 3.27*** | 3.185*** | 2.911*** | 3.348*** | 29.186*** |
| Research performance | $X_{11}$ | Tot. publications | 1.037*** | 1 | 1.051*** | 1.066*** | 1.045*** | 1.029*** | 1.082*** | 1.058*** | 1.042*** | 1.049*** |
| | $X_{12}$ | FSS | 0.906 | 1.144* | 0.721*** | 0.843 | 0.886*** | 0.924*** | 0.805*** | 0.795** | 0.888** | 1.01 |
| | $X_{13}$ | Top 20% | 0.868 | 1.199 | 1.029 | 0.844 | 1.265** | 1.132 | 0.878 | 1.029 | 0.929 | 0.831 |
| Environment | $X_{14}$ | Peers collaborating | 2.901*** | 2.607*** | 2.867*** | 3.247*** | 2.315*** | 2.387*** | 2.373*** | 3.847*** | 2.764*** | 1.637 |
| University type | $X_{15}$ | Private | 0.532 | 1.183 | 1.278 | omitted | 0.617* | 0.687*** | 0.715 | 0.565 | 2.003** | 1.277 |
| | $X_{16}$ | Polytechnic or SS | 1.029 | 0.996 | 0.878 | 1.871 | 1.215 | 1.54 | 0.542 | 1.111 | 1.163 | omitted |
| University size | $X_{17}$ | Medium | 0.924 | 0.772 | 0.681** | 1.087 | 0.957 | 0.743** | 0.913 | 0.48*** | 1.453** | 0.251** |
| | $X_{18}$ | Large | 0.789 | 0.817 | 0.544*** | 0.76 | 0.757** | 0.654*** | 0.802 | 0.51*** | 1.219 | 0.407* |
| University location | $X_{19}$ | South | 1.191 | 0.603** | 1.448** | 1.012 | 1.069 | 1.521*** | 1.218 | 1.66 | 1.429** | 0.431 |
| | $X_{20}$ | Center | 0.811 | 0.752 | 1.43** | 1.212 | 1.04 | 1.406*** | 2.272*** | 2.554*** | 1.459*** | 0.363* |
| | $X_{21}$ | Northeast | 1.073 | 0.753 | 1.522*** | 1.243 | 1.165 | 1.516*** | 2.338*** | 2.303** | 1.451*** | 0.389* |
| | $X_{22}$ | Northwest | 1.064 | 0.842 | 2.177*** | 1.11 | 1.284* | 1.553*** | 2.204*** | 3.004*** | 1.462** | 0.923 |
| | | Obs. | 2793 | 2118 | 2808 | 996 | 4605 | 8871 | 2857 | 1402 | 5081 | 1261 |
| | | Pseudo $R^2$ | 0.195 | 0.111 | 0.150 | 0.148 | 0.127 | 0.158 | 0.165 | 0.175 | 0.151 | 0.224 |

*§ 1, Mathematics and computer science; 2, Physics; 3, Chemistry; 4, Earth sciences; 5, Biology; 6, Medicine; 7, Agriculture and veterinary sciences; 8, Civil engineering; 9, Industrial and information engineering; 10, Psychology*

*Statistical significance: * p-value <0.10, ** p-value <0.05, *** p-value <0.01.*

## 5. Conclusions

Over the past four decades, governments, industry, and funding organizations have increasingly pressured universities to contribute to national innovation processes and to become more “entrepreneurial” (Todorovic, McNaughton, & Guild, 2011; Etzkowitz, 1983). Alongside their more traditional functions of research and teaching, they are now expected to fulfill the so-called “third role” (Etzkowitz & Leydesdorff, 1995). One of the means for universities to pursue this new mission is to engage in research collaboration with industry. The potentials of this strategy have gained increasing attention among scholars and action on the part of policy-makers in the form of incentivizing initiatives (Davenport, Davies, & Grimes, 1998; Debackere & Veugelers, 2005).

From the universities’ viewpoint, the economic incentives embedded in industry collaboration (e.g. access to financial resources, complementary material assets) are not always enough to support the engagement desired. Countering the attractions are disincentives, in particular transactions costs (Belkhodja & Landry, 2005; Drejer & Jørgensen, 2005), which will clearly vary with the level of heterogeneity among members of the academic-industry research team (Abramo, D’Angelo, Di Costa, & Solazzi, 2011).

Furthermore, the implementation of national performance-based research funding systems can in fact negatively affect the policies incentivizing public-private collaborations. It is known that academics’ collaboration with colleagues in the private sector leads to publications with average impact lower than those stemming from intra-sector collaborations, and to a lower proportion of highly-cited publications. The same holds true when foreign co-authors are involved (Abramo, D’Angelo, & Di Costa, 2020). From the perspective of the public scientist, increasingly subject to assessments based on the scholarly impact of their scientific activity, this awareness might act as a deterrent. This leads to counter-purposes, or at least trade-offs, between policies aimed at increasing public researcher productivity and others aimed at encouraging involvement in industrial research collaborations.

For all these reasons, policy-makers, industry and the university managers would clearly benefit from understanding the main drivers of academic engagement in collaboration with industry. With better knowledge of individual and contextual features, incentive systems can be better formulated and targeted. For just and effective operation, the relevant variables should also be taken into account in performance measurements and reward systems. Considering only one example: to demand that the academics of Engineering and Mathematics, or those of northern vs southern Italy, all achieve the same intensity of industrial research collaboration would be unfair, as well as dysfunctional. The empirical results of the current study, in fact, illustrate the varying cross-sector collaboration propensities of research fields, aligning with findings by Tijssen (2012). Considering the potentials of Italian policy in particular, they show the outstanding effect of territorial location on academics, reflecting the disproportionate concentration of industry in the north vs south and islands.

Other relevant results of the analysis are that male rather than female scientists show greater propensity to interact with industry. Different explanations come to mind, with implications for academic policy, social policy, and management. Women could be more resistant to territorial movement for reasons of simultaneous involvement in other roles, particularly in family. Throughout the world, there remains a cultural heritage in which industry (in particular manufacturing, with which Engineering academics tend to collaborate most) is seen as mainly for men (EU, 2017; Abramo, D’Angelo, & Murgia,

2013b).

We also find, all other factors being equal, that it is the professor under age 40 who engages most in research collaboration with industry; it is the full professor, the top performer, the one with highly diversified research, and who has a certain tradition in collaborating with industry. This professor is likely to be part of a staff used to collaborating with industry, in a small university, typically a polytechnic or a school for advanced studies. All of this presents a profile very similar to that extrapolated by Perkmann, Salandra, Tartari, McKelvey, and Hughes (2021), in their recent review of the literature on the theme.

The findings of empirical investigations such as ours can be generalized to other countries with extreme caution. For some results in particular, comparability to any other countries would require similarity in the contextual variables: cultural, social, industrial, and of the research systems themselves.

Moreover, the intrinsic limits of all analyses adopting bibliometric techniques require caution. First of all, the circumstances that not all universities-industry research collaborations lead to results encodable in scientific publications, and not all university/industry co-authored publications are the outcome of real research collaborations. Also, the bibliographic repertories (such as WoS, used here) do not register all publications. Furthermore, the measurement of research performance by citation-based indicators (FSS) is subject to a number of limits and assumptions. Finally, citations can also be negative or inappropriate, and in any case they certify only scholarly impact, forgoing other types of impact. These limitations should always be recalled when interpreting findings arising from bibliometric techniques.

The current study has approached the issue of academic-industry collaboration from the view of the academics. It would be equally interesting to investigate from the industry perspective, concerning the drivers of engaging in private-public collaboration.

## References

Abramo, G., & D'Angelo, C.A. (2014). How do you define and measure research productivity? *Scientometrics*, 101(2), 1129-1144. DOI: 10.1007/s11192-014-1269-8

Abramo, G., D'Angelo, C. A., & Murgia, G. (2013a). The collaboration behaviours of scientists in Italy: A field level analysis. *Journal of Informetrics*, *7*(2), 442–454. DOI: 10.1016/j.joi.2013.01.009

Abramo, G., D'Angelo, C.A., & Di Costa, F. (2020). The relative impact of private research on scientific advancement. Working paper, http://arxiv.org/abs/2012.04908, last accessed on 17 December 2020.

Abramo, G., D'Angelo, C.A., & Murgia, G. (2013b). Gender differences in research collaboration. *Journal of Informetrics*, 7(4), 811-822. DOI: 10.1016/j.joi.2013.07.002

Abramo, G., D'Angelo, C.A., & Murgia, G. (2014). Variation in research collaboration patterns across academic ranks. *Scientometrics*, 98(3), 2275-2294. DOI: 10.1007/s11192-013-1185-3

Abramo, G., D'Angelo, C.A., & Murgia, G. (2017). The relationship among research productivity, research collaboration, and their determinants. *Journal of Informetrics*, 11(4), 1016-1030. DOI: 10.1016/j.joi.2017.09.007

Abramo, G., D'Angelo, C.A., Di Costa, F., & Solazzi, M. (2011). The role of information asymmetry in the market for university-industry research collaboration. *The Journal of Technology Transfer*, 36(1), 84-100. DOI: 10.1007/s10961-009-9131-5.

Balconi, M., & Laboranti, A. (2006). University-industry interactions in applied research: The case of microelectronics. *Research Policy*, 35(10), 1616-1630. DOI: 10.1016/j.respol.2006.09.018

Bayer, A. E., & Smart, J. C. (1991). Career publication patterns and collaborative "styles" in American academic science. *Journal of Higher Education*, *62*(6), 613–636.

Bekkers, R., & Bodas Freitas, I. M. (2008). Analysing knowledge transfer channels between universities and industry: To what degree do sectors also matter? *Research Policy*, *37*(10), 1837-1853. DOI: 10.1016/j.respol.2008.07.007

Belkhodja, O., & Landry, R. (2005). The Triple Helix collaboration: Why do researchers collaborate with industry and the government? What are the factors influencing the perceived barriers? *5th Triple Helix Conference*, 1–48.

Berbegal-Mirabent, J., Sánchez García, J. L., & Ribeiro-Soriano, D. E. (2015). University-industry partnerships for the provision of R&D services. *Journal of Business Research*, *68*(7), 1407-1413. DOI: 10.1016/j.jbusres.2015.01.023

Bercovitz, J., & Feldman, M. (2003). Technology transfer and the academic department: who participates and why? In: Paper presented at the *DRUID Summer Conference 2003*, Copenhagen, June 12–14.

Blumenthal, D., Campbell, E. G., Causino, N., & Louis, K. S. (1996). Participation of life science faculty in research relationships with industry. *New England Journal of Medicine*, *335*(23), 1734-1739. DOI:10.1056/NEJM199612053352305

Boschini, A., & Sjögren, A. (2007). Is team formation gender neutral? Evidence from coauthorship patterns. *Journal of Labor Economics*, *25*(2), 325–365. DOI: 10.1086/510764

Bowyer, K., Kranenburg, C., & Dougherty, S. (2001). Edge detector evaluation using empirical ROC curves. *Computer Vision and Image Understanding*, *84*(1), 77-103.

Bozeman, B., & Corley, E. (2004). Scientists' collaboration strategies: Implications for scientific and technical human capital. *Research Policy*, 33(4), 599–616. DOI: 10.1016/j.respol.2004.01.008

Bozeman, B., & Gaughan, M. (2011). How do men and women differ in research collaborations? An analysis of the collaborative motives and strategies of academic researchers. *Research Policy*, *40*(10), 1393–1402. DOI: 10.1016/j.respol.2011.07.002

Calvo, N., Fernández-López, S., & Rodeiro-Pazos, D. (2019). Is university-industry collaboration biased by sex criteria? *Knowledge Management Research and Practice*, *17*(4), 408-420. DOI: 10.1080/14778238.2018.1557024

D'Angelo, C. A., Giuffrida, C., & Abramo, G. (2011). A heuristic approach to author name disambiguation in large-scale bibliometric databases. *Journal of the American Society for Information Science and Technology*, 62(2), 257-69.

D'Este, P., & Patel, P. (2007). University-industry linkages in the UK: What are the factors underlying the variety of interactions with industry? *Research Policy*, *36*(9), 1295–1313. DOI: 10.1016/j.respol.2007.05.002

Davenport, S., Davies, J., & Grimes, C. (1998). Collaborative research programmes: building trust from difference. *Technovation*, *19*(1), 31–40. DOI: 10.1016/S0166-4972(98)00083-2.

Debackere, K., & Veugelers, R. (2005). The role of academic technology transfer organizations in improving industry science links. *Research Policy, 34*(3), 321–342. https://doi.org/10.1016/j.respol.2004.12.003

Di Gregorio, D., Shane, S., 2003. Why do some universities generate more start-ups than others? *Research Policy*, 32(2), 209–227

Drejer, I., & Jørgensen, B. H. (2005). The dynamic creation of knowledge: Analysing public-private collaborations. *Technovation*, *25*(2), 83–94. DOI: 10.1016/S0166-4972(03)00075-0

Etzkowitz, H. (1983). Entrepreneurial scientists and entrepreneurial universities in American academic science. *Minerva*, *21*(2–3), 198–233. https://doi.org/10.1007/BF01097964.

Etzkowitz, H., & Leydesdorff, L. (1995). The Triple Helix: University - Industry - Government Relations A Laboratory for Knowledge Based Economic Development. *EASST Review*, *14*, 14–19.

EU (2017). Special Eurobarometer 465: Gender Equality 2017. *EU Open Data Portal*, https://data.europa.eu/euodp/en/data/dataset/S2154_87_4_465_ENG, last accessed on 17 December 2020.

Fan, X., Yang, X., & Chen, L. (2015). Diversified resources and academic influence: patterns of university–industry collaboration in Chinese research-oriented universities. *Scientometrics*, *104*(2), 489–509. DOI: 10.1007/s11192-015-1618-2

Garcia, R., Araújo, V., Mascarini, S., Santos, E. G., & Costa, A. R. (2020). How long-term university-industry collaboration shapes the academic productivity of research groups. *Innovation: Organization and Management*, *22*(1), 56-70. DOI: 10.1080/14479338.2019.1632711

Giuri, P., Munari, F., Scandura, A., & Toschi, L. (2019). The strategic orientation of universities in knowledge transfer activities. *Technological Forecasting and Social Change*, *138*, 261-278. DOI: 10.1016/j.techfore.2018.09.030

Guimón, J., & Paunov, C. (2019). Science-industry knowledge exchange: A mapping of policy instruments and their interactions. *OECD Science, Technology and Industry Policy Papers*, No. 66, OECD Publishing, Paris, DOI: 10.1787/66a3bd38-en, last accessed on 17 December 2020.

He, Z., Geng, X., & Campbell-Hunt, C. (2009). Research collaboration and research output: A longitudinal study of 65 biomedical scientists in a New Zealand university. *Research Policy*, *38*(2), 306-317. DOI: 10.1016/j.respol.2008.11.011

Huang, M.-H., Yang, H.-W., & Chen, D.-Z. (2015). Industry–academia collaboration in fuel cells: a perspective from paper and patent analysis. *Scientometrics*, *105*(2), 1301–1318. DOI: 10.1007/s11192-015-1748-6

Iorio, R., Labory, S., & Rentocchini, F. (2016). The importance of pro-social behaviour for the breadth and depth of knowledge transfer activities: An analysis of italian academic scientists. *Research Policy*, *46*(2), 497-509. DOI: 10.1016/j.respol.2016.12.003

Katz, J. S., & Martin, B. R. (1997). What is research collaboration? *Research Policy*, *26*(1), 1–18. DOI: 10.1016/S0048-7333(96)00917-1

Lee, S., & Bozeman, B. (2005). The impact of research collaboration on scientific productivity. *Social Studies of Science*, *35*(5), 673-702. DOI: 10.1177/0306312705052359

Llopis, O., Sánchez-Barrioluengo, M., Olmos-Peñuela, J., & Castro-Martínez, E. (2018). Scientists’ engagement in knowledge transfer and exchange: Individual factors, variety of mechanisms and users. *Science and Public Policy*, *45*(6), 790-803. DOI:

10.1093/scipol/scy020
Mansfield, E. (1995). Academic research underlying industrial innovations: sources, characteristics, and financing. *Review of Economics and Statistics*, 77(1), 55-65
Mansfield, E., & Lee, J.Y. (1996). The modern university: contributor to industrial innovation and recipient of industrial R&D support. *Research Policy*, 25 (7), 1047-1058
Mowery, D. C., Nelson, R. R., Sampat, B. N., & Ziedonis, A. A. (2001). The growth of patenting and licensing by U.S. universities: An assessment of the effects of the Bayh-Dole act of 1980. *Research Policy*, *30*(1), 99–119. DOI: 10.1016/S0048-7333(99)00100-6
Muscio, A, & Pozzali, A. (2013). The effects of cultural distance in university-industry collaborations: some evidence from Italian universities. *Journal of Technology Transfer*, 38(4), 486-508
Perkmann, M., King, Z., & Pavelin, S. (2011). Engaging excellence? effects of faculty quality on university engagement with industry. *Research Policy*, *40*(4), 539-552. DOI: 10.1016/j.respol.2011.01.007
Perkmann, M., Salandra, R., Tartari, V., McKelvey, M., & Hughes, A. (2021). Academic engagement: A review of the literature 2011-2019. *Research Policy*, *50*(1) DOI:10.1016/j.respol.2020.104114
Perkmann, M., Tartari, V., McKelvey, M., Autio, E., Broström, A., D'Este, P., . . . Sobrero, M. (2013). Academic engagement and commercialisation: A review of the literature on university-industry relations. *Research Policy*, *42*(2), 423-442. DOI: 10.1016/j.respol.2012.09.007
Phan, P H, & Siegel, D S. (2006). The effectiveness of university technology transfer. *Foundations and Trends in Entrepreneurship*, 2(2), 77-144
Schartinger, D., Schibany, A., & Gassler, H. (2001). Interactive relations between university and firms: empirical evidence for Austria. *Journal of Technology Transfer*, 26, 255–268.
Shane, S. (2004). Encouraging university entrepreneurship? The effect of the Bayh-Dole Act on university patenting in the United States. *Journal of Business Venturing*, *19*(1), 127–151. DOI: 10.1016/S0883-9026(02)00114-3
Tartari, V., & Salter, A. (2015). The engagement gap: Exploring gender differences in university - industry collaboration activities. *Research Policy*, *44*(6), 1176-1191. DOI: 10.1016/j.respol.2015.01.014
Tartari, V., Perkmann, M., & Salter, A. (2014). In good company: The influence of peers on industry engagement by academic scientists. *Research Policy*, *43*(7), 1189-1203. DOI:10.1016/j.respol.2014.02.003
Thune, T., Reymert, I., Gulbrandsen, M., & Aamodt, P. O. (2016). Universities and external engagement activities: Particular profiles for particular universities? *Science and Public Policy*, *43*(6), 774-786. DOI: 10.1093/scipol/scw01.
Thursby, J. G., & Thursby, M. C. (2003). University Licensing and the Bayh-Dole Act. *Science*, *301*(5636), 1052–1052.
Tijssen, R. J. W. (2012). Co-authored research publications and strategic analysis of public-private collaboration. *Research Evaluation*, *21*(3), 204–215. https://doi.org/10.1093/reseval/rvs013
Todorovic, Z. W., McNaughton, R. B., & Guild, P. (2011). ENTRE-U: An entrepreneurial orientation scale for universities. *Technovation*, *31*(2–3), 128–137. https://doi.org/10.1016/j.technovation.2010.10.009

Traoré, N., & Landry, R. (1997). On the determinants of scientists' collaboration. *Science Communication*, *19*(2), 124–140.

Ubfal, D., & Maffioli, A. (2011). The impact of funding on research collaboration: Evidence from a developing country. *Research Policy*, *40*(9), 1269–1279. DOI: 10.1016/j.respol.2011.05.023

Weerasinghe, I. M. S., & Dedunu, H. H. (2020). Contribution of academics to university–industry knowledge exchange: A study of open *innovation* in Sri Lankan universities. *Industry and Higher Education*, DOI: 10.1177/0950422220964363

Zhao, Z., Broström, A., & Cai, J. (2020). Promoting academic engagement: University context and individual characteristics. *Journal of Technology Transfer*, *45*(1), 304-337. DOI: 10.1007/s10961-018-9680-6